\documentclass[aps,prc,twocolumn,showpacs,groupedaddress,floatfix]
{revtex4}

\usepackage{graphicx}
\begin{document}


\title{Application of the Pauli principle in many-body scattering}


\author{S.~P.~Weppner}
\email[]{weppnesp@eckerd.edu}
\affiliation{Natural Sciences, Eckerd College, St. Petersburg, 
Florida 33711}


\date{\today}

\begin{abstract}
A new development in the
antisymmetrization of the first-order nucleon-nucleus elastic
microscopic optical
potential
is presented which systematically
includes the many-body character of the
nucleus within the two-body scattering operators. 
The results reduce the overall strength of the
nucleon-nucleus potential and require the inclusion
of
historically excluded channels from the nucleon-nucleon
potential input.
Calculations produced
improve the match with neutron-nucleus total 
cross section, elastic proton-nucleus differential cross section,
and spin observable data. A comparison is also done using
different nucleon-nucleon potentials from the past twenty years.
\end{abstract}


\maketitle


\section{Introduction}\label{2B}


The incorporation of antisymmetrization, or the Pauli principle,
into the microscopic many-body scattering
problem has
been a subject of study
for five decades. The underlying problem is developing a
many-body scattering theory  which uses the
two-body nucleon-nucleon interaction while
still treating all particles as indistinguishable fermions.
The pioneering work of Watson and
collaborators~\cite {watson2,watson3} included
the Pauli principle by antisymmetrizing only the
active two-body projectile-target nucleon interaction. 

In the late 1970's  and early 1980's there
was a renewed interest in the scattering
theory of indistinguishable particles~\cite{kowalski-rev}.
These studies brought a new level of mathematical sophistication to
the subject of many-body antisymmetrization
but had little effect on practical
calculations. Some of these theoretical developments
were rigorous and  complete~\cite{kowalski1,kowalski4}. 
They began with a well defined
many-body microscopic 
optical potential formulated using the connected kernel and
unitary properties of Faddeev~\cite{fadeev1} and Alt, 
Grassberger, and Sandhas~\cite{AGS}, however, 
because of their complexity,
only results
for the lightest of nuclei were possible. When  
simplified to a tractable
problem they were reduced back to the two-body Watson  
approximation~\cite{kozack2,kowalski2}.
Others during the same time period started with the Watson 
theory
and then included higher order terms in a consistent 
manner~\cite{bolle, pickle1} using a cluster or spectator 
expansion. These terms were also complex and had limited
use (see Ref.~\cite{ray1}). 

This work advances an alternative approach 
which modifies the Watson
approximation simply and is therefore useful in calculation.
The Pauli principle is treated in a simple many-body
representation
which is practical for all microscopic optical
potential calculations
which
use the nucleon-nucleon potential and a nuclear structure
calculation as inputs. 
In section~\ref{sec3.1} and section~\ref{sec3.2}
a brief presentation of
the theory of microscopic optical potentials is given. 
A discussion of the distinguishability of the projectile in 
nucleon-nucleus scattering and a simple modification is 
presented in section~\ref{CM}. A better  modification is
developed in section~\ref{MB}. Comparisons between the different
formulations of antisymmetrization and experiment
are made in section~\ref{sec3.6}.
In section~\ref{sec3.8}
a summary is given.

\section{Background}\label{sec3.1}
It is customary in a microscopic
formulation of nucleon-nucleus scattering
for the external interaction between the two fragments
to be defined as the sum of all
nucleon-nucleon interactions, $V_{0i}$, between the external
projectile nucleon
labeled `0' and  the  internal nucleons of the $A$ body target:
\begin{equation}
V=\sum_{i=1}^A V_{0i}. \label{eq4.4}
\end{equation}
By following the methods of
Watson~\cite{watson2,watson3}
and Feshbach~\cite{feshbach2}
we may define the optical potential by splitting
Hilbert space into two orthogonal projections.
The projections ${\cal P}$
and ${\cal Q}$ define the
elastic and inelastic scattering projections respectively.
Using these projections we may
define the many-body transition operator $T$ as
\begin{equation}
T=U+U{\cal G}_0{\cal P}T, \label{eq4.7}
\end{equation}
where $U$ is the optical potential defined as
\begin{equation}
U=V+V{\cal G}_0{\cal Q}U,  \label{eq4.8}
\end{equation}
and ${\cal G}_0$ is the many-body propagator.
These projections divides elastic scattering
into two parts. While calculating the optical potential $U$,
using Eq.~(\ref{eq4.8}), the projection ${\cal Q}$ constrains the
intermediate state interactions to take place in the 
excited energy region of the nuclear target~\cite{bolle}.

\begin{figure}
\includegraphics[width=8.5cm]{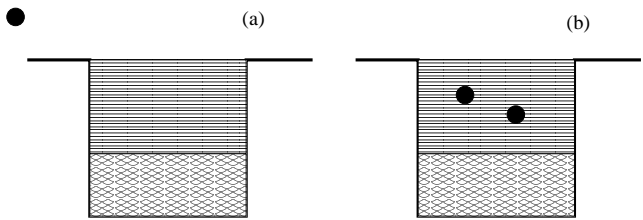}
\caption{(a) Nucleon-nucleus scattering before the interaction
begins. (b) Nucleon-nucleus first-order scattering. The black
circles represent nucleons. The well represents the
potential well of the nucleus. The
ground state levels of the
nuclear well are  represented by a cross-hatch and the excited
states are represented by horizontal lines. 
The interaction takes place
in the excited state 
continuum with only one target nucleon from the nucleus
interacting with the projectile.\label{fig4.1}}
\end{figure}
The first-order
approximation to Eq.~(\ref{eq4.8}) which is valid
at suitably high energies
is to assume that the projectile only interacts with one
target nucleon in the nucleus~\cite{watson2}.
Applying this assumption
we have 
\begin{equation}
U=\sum_{i=1}^AU_{i}\approx
\sum_{i=1}^A \hat{\tau_{0i}}, 
\end{equation}
where $\hat{\tau_{0i}}$ is
defined from Eq.~(\ref{eq4.8}) as
\begin{equation}
\hat{\tau_{0i}}=V_{0i}+
V_{0i}{\cal G}_0{\cal Q}\hat{\tau_{0i}}, \label{eq4.9}
\end{equation}
where the index `i' represents  a specific nucleon in the target.
The product 
function ${\cal G}_0{\cal Q}$ is a many-body operator product,
which may include effects from
of the binding energy, nuclear medium, intermediate state
Pauli blocking,  and
excited states of the nucleus.
To make the
problem tractable these operators are reduced to
a two-body operator $({\cal G}_0{\cal Q})_{2B}$
by various approximations such that
\begin{equation}
U_i\approx{\tau}_{0i}=V_{0i}
+V_{0i}({\cal G}_0{\cal Q})_{2B}{\tau}_{0i} \label{eq4.10}
\end{equation}
can be calculated~\cite{ray1,amos1}.

In Fig.~\ref{fig4.1} we depict an interaction between the
projectile and single target nucleon.
Before the interaction begins the nucleus is in its ground
state 
but as the collision proceeds
the projectile
enters the nuclear excited states  and
interacts with one bound nucleon. 
The $A-1$ non-active nucleons are
labeled the spectator core because they play no direct
role in the two-body interaction~\cite{bolle,friedman}.

To calculate the elastic nucleon-nucleus optical potential
a summation of Eq.~(\ref{eq4.10}) over all 
target nucleons is required  followed by a projection
onto the elastic channel 
\begin{equation}
{\cal P}U{\cal P}=
\langle \Phi_A|\;\langle {\bf{k'_0,k'_i}}|\;
\sum_{i=1}^A{\tau}_{0i}
\; |{\bf{k_0,k_i}}\rangle\;
|\Phi_A\rangle,  \label{eq4.11}
\end{equation}
where $|\bf{k_0}\rangle$ is the momentum of the
projectile nucleon with
label `0'
and $|\Phi_A\rangle$ is the ground state basis of the
target nucleus. For simplicity the spin and isospin characteristics 
of the nucleons have not been explicitly included in 
Eq.~(\ref{eq4.11}), but they will be discussed later.

The wave function of the target nucleus
is usually in a convenient single particle basis
used to define the nuclear structure density, $\rho_A$, as
\begin{equation}
\rho_A({\bf {\tilde{k}'_i}},{\bf \tilde{k}_i})\equiv
\langle \Phi_A'|{\bf {\tilde{k}'_i}}\rangle
\;\langle {\bf \tilde{k}_i}|\Phi_A\rangle, \label{eq4.13}
\end{equation}
where ${\bf \tilde{k}_i}$ measures the momentum of target
nucleon `i' from the
center of the struck nucleus.
The definition of the optical potential
is thus reduced to the
traditional $t\rho$ form developed by Watson~\cite{watson2}
and Kerman, McManus,
and Thaler~\cite{KMT}
\begin{eqnarray}
{\cal P}U{\cal P}\equiv \sum_{i=1}^A\;
\int &&d{\bf {\tilde{k}'_i}}\; d{\bf {\tilde{k}_i}}\;
\langle {\bf{k'_0,k'_i}}|{\tau}_{0i}
|{\bf{k_0,k_i}}\rangle
\rho_A({\bf {\tilde{k}'_i}},{\bf \tilde{k}_i}) \nonumber \\
&& \delta(\bf{k'_0}+
{\bf {{k}'_i}}-\bf{k_0}-{\bf {{k}_i}}),
\label{eq4.14}
\end{eqnarray}
where the delta function conserves momentum 
for the projectile-target nucleon system.
The elastic 
nucleon-nucleus first-order optical potential can therefore
be calculated  using a nuclear
density structure calculation and a two-body interaction, $\tau_{0i}$,
which contains the bare nucleon-nucleon potential $V_{0i}$
as defined by Eq.~(\ref{eq4.9}). These
two inputs are the foundation  of every microscopic 
nucleon-nucleus optical potential
calculation.

In this work we will examine the antisymmetric 
character of this two-body
interaction including the spin and isospin 
dependencies
in the context of the traditional Watson approximation.
When discussing the two-body interaction a
relationship between the two-nucleon
state and the scattering state basis may be defined as
\begin{widetext}
\begin{eqnarray}
&&|\Psi_{2B}\rangle=|{\bf q},{\bf K}\rangle|S,m_s\rangle|T,m_t\rangle
=\sum_{{m_s}_0,{m_t}_0}
\biggl|{\bf k_0},{m_s}_0,{m_t}_0,
\;{\bf k_i},{m_s}_i,{m_t}_i\biggr\rangle
\biggl\langle {m_s}_0,{m_t}_0,\;
{m_s}_i,{m_t}_i\biggm |
S,m_s,T,m_t\biggr\rangle
\label{eq14}
\end{eqnarray}
\end{widetext}
where
$m_s={m_s}_0+{m_s}_i, m_t={m_t}_0+{m_t}_i$,
${\bf K} = {{\bf k_i}+{\bf k_0}}$,
${\bf q} = {\bf k_i}-{\bf k_0}$,
and the identity of the 
particle is represented by the subscript
`0' or `$i$'. This defines the
traditional two nucleon state which has by definition of the 
individual nucleons
${m_s}_i=\pm\frac{1}{2}$ (spin up or down)
and ${m_t}_i=\pm\frac{1}{2}$ (proton or neutron). 
This is a
basis which the $\tau_{0i}$ operator of Eq.~(\ref{eq4.11}) and
Eq.~(\ref{eq4.14}) is typically calculated in.

According to the Pauli exclusion principle, since the 
nucleons are fermions, the
scattering state 
$|{\bf q},{\bf K}\rangle|S,m_s\rangle|T,m_t\rangle_{2B}$,
described by Eq.~(\ref{eq14}), must be antisymmetric. The possible
choices are: \\
\begin{eqnarray}
&&|{\bf 1}\rangle\equiv 
|{\bf q},{\bf K}\rangle_{sym}|S=0,m_s\rangle|T=1,m_t\rangle\nonumber\\
&&|{\bf 2}\rangle\equiv
|{\bf q},{\bf K}\rangle_{asym}|S=1,m_s\rangle|T=1,m_t\rangle\nonumber\\
&&|{\bf 3}\rangle\equiv
|{\bf q},{\bf K}\rangle_{sym}|S=1,m_s\rangle|T=0,m_t\rangle\nonumber\\
&&|{\bf 4}\rangle\equiv
|{\bf q},{\bf K}\rangle_{asym}|S=0,m_s\rangle|T=0,m_t\rangle.
\label{list1}
\end{eqnarray}
These states are traditionally  described in either a partial wave or
helicity basis. Explicitly, for neutron-proton 
scattering, all four states
are included. In proton-proton scattering only $T=1$ 
is possible so states $|{\bf 3}\rangle$ and $|{\bf 4}\rangle$
are excluded, 
however because of the exact identical nature 
of the two protons, the identity of the original nucleons is lost during
the scattering process so the contributed strengths of 
states $|{\bf 1}\rangle$ and $|{\bf 2}\rangle$
are doubled. This is the traditional kinematical
doubling for identical particles in quantum mechanics. For example in
the two-body center of mass frame a scattering
of 10$^o$ is equivalent to a scattering of 170$^o$ because the two
protons cannot be differentiated thereby doubling the strength at 
both angles.

These results are well known but reiterated here because 
later we will
change the character of this two-body basis 
to suit the many-body problem.
For future reference we define equivalent to Eq.~(\ref{eq14}) and
Eq.~(\ref{list1})
\begin{eqnarray}
|\Psi_{2B}\rangle_{asym}&&\equiv
|{\bf q},{\bf K}\rangle|S,m_s\rangle|T,m_t\rangle_{asym} 
\nonumber \\ 
&&\mbox{and in practice} \nonumber \\
|\Psi_{2B-np}\rangle&&\equiv
\frac{|{\bf 1}\rangle+|{\bf 2}\rangle+|{\bf 3}\rangle+
|{\bf 4}\rangle}{2} \nonumber \\
|\Psi_{2B-pp}\rangle&&\equiv
\frac{\sqrt{2}|{\bf 1}\rangle+\sqrt{2}|{\bf 2}\rangle}{2},
\label{eq4.11a}
\end{eqnarray}
using the states as defined in Eq.~(\ref{list1}). The $2B-np$ describes
the traditional neutron-proton antisymmetric basis state while $2B-pp$
describes either the proton-proton or neutron-neutron basis states. To
calculate a proton-$^{16}$O optical potential 
for example we would sum up
eight $\tau_{0i}$ interactions in the traditional 
proton-proton basis and 
eight $\tau_{0i}$ interactions in the traditional 
neutron-proton basis because
the proton projectile  interacts with either one of eight 
protons or eight neutrons in the target nucleus. They are also
weighted with the
appropriate single particle $^{16}$O density contribution
\begin{eqnarray}
&&{\cal P}U{\cal P}\equiv\nonumber \\
&8&\int d{\bf {\tilde{k}'_i}}\; d{\bf {\tilde{k}_i}}\;
\langle \Psi_{2B-pp} |{\tau}_{0i}
|\Psi_{2B-pp}\rangle
\rho_{proton}({\bf {\tilde{k}'_i}},{\bf \tilde{k}_i}) \nonumber \\
+&8&\int d{\bf {\tilde{k}'_i}}\; d{\bf {\tilde{k}_i}}\;
\langle \Psi_{2B-np} |{\tau}_{0i}
|\Psi_{2B-np}\rangle
\rho_{neutron}({\bf {\tilde{k}'_i}},{\bf \tilde{k}_i}),\nonumber\\ 
\label{eq4.14b}
\end{eqnarray}
akin to Eq.~(\ref{eq4.14}).
The validity of
this calculation in the context of the 
many-body problem will be examined in section~\ref{sec3.3}.

\section{Practical first-order 
many-body calculations}\label{sec3.2}
The two-body interaction calculated in many microscopic
nucleon-nucleus
interactions is Eq.~(\ref{eq4.9})
\begin{eqnarray*}
\hat{\tau_{0i}}=V_{0i}+V_{0i}{\cal G}_0{\cal Q}\hat{\tau_{0i}}.
\end{eqnarray*}
As stated previously, ${\cal G}_0{\cal Q}$ is
a product of many-body operators which
involve the propagation of two nucleons through a
nucleus which is not in the ground state.
The spectator core
directly affects the two nucleons' propagation, identity, and
interaction. In this section we will discuss the two
most common methods which make this operator suitable for
calculation. 
For a more exhaustive summary of these two techniques see
Ref.~\cite{amos1}.

\subsection{Folding $t$ operator approach}\label{ssect}

The  first-order optical potential
involves a sum of two-body interactions between the
projectile and the target nucleons
\begin{equation}
U = \sum_{i=1}^{A}\hat{\tau_{0i}} , \label{eq:2.10}
\end{equation}
where the operator $\hat{\tau_{0i}}$ is
\begin{eqnarray}
\hat{\tau_{0i}} &=& V_{0i} + V_{0i} {\cal G}_0 {\cal Q}
\hat{\tau_{0i}} \nonumber \\
&=& V_{0i} + V_{0i}{\cal G}_0
\hat{\tau_{0i}} - V_{0i}{\cal G}_0 {\cal P} \hat{\tau_{0i}}\nonumber \\
&=& {\cal T}_{0i} - {\cal T}_{0i} {\cal G}_0
{\cal P} \hat{\tau_{0i}}, \label{eq:2.11}
\end{eqnarray}
which is derived by 
using the relationship ${\cal P} +{\cal Q}\equiv  1$.
This procedure successfully
removes operator ${\cal Q}$ from the
calculation with the cost of a new operator ${\cal T}_{0i}$.

For elastic scattering only
${\cal P}\hat{\tau_{0i}}{\cal P}$ 
need to be 
considered.  Explicitly it appears as
\begin{equation}
\hat{\tau_{0i}} = 
{\cal T}_{0i}
-{\cal T}_{0i}
\frac {1}
{(E-E_A) - h_0 + i\varepsilon}\hat{\tau_{0i}},
 \label{eq:2.12}
\end{equation}
where ${\cal T}_{0i}$
is defined as the solution of the
sum of nucleon-nucleon interactions as the two nucleons
propagate through
the many-body medium
\begin{equation}
{\cal T}_{0i} = V_{0i} + V_{0i}
{\cal G}_0 {\cal T}_{0i}. \label{eq:2.13}
\end{equation}

Since Eq.~(\ref{eq:2.12}) is a simple two-body integral equation,
the principal problem is to find a solution of Eq.~(\ref{eq:2.13}),
which still has a  many-body character due to the propagator
\begin{equation}
{\cal G}_0=(E -h_{0} -H_{A} +i\varepsilon)^{-1},
\end{equation}
where
$H_A$ is the many-body Hamiltonian of the target nucleus.
If the propagator ${\cal G}_{0}$
is expanded
within a single particle description,
one obtains to  first-order \cite{med2,med1,density}
\begin{equation}
G_i = [(E-E^i) -h_0 -h_i -W_i + i\varepsilon]^{-1}, \label{eq:2.14}
\end{equation}
where $h_i$ is
the kinetic energy of the $i$th target particle and
$W_i=\sum_{j\neq i}v_{ij}$.
The quantity $W_i$ represents the mean field acting between the struck
target nucleon and the residual (A-1) nucleus made from summing up all
the individual nucleon-nucleon potentials contained within the 
spectator core.
The operator
$ {\cal T}_{0i}$  of Eq.(\ref{eq:2.13}) is then approximated as
\begin{eqnarray}
{\cal T}_{0i}&\approx&=
V_{0i} + V_{0i} G_i\tau_{0i} \nonumber \\
  &=& {\cal T}_{2B} + {\cal T}_{2B} G_0 W_i G_i
{\cal\tau}_{0i}, \label{eq2:2.15}
\end{eqnarray}
where the pure nucleon-nucleon interaction 
operator ${\cal T}_{2B}$ is defined
as
\begin{equation}
{\cal T}_{2B} = V_{0i} + V_{0i} G_0 {\cal T}_{2B}, \label{eq:2.17}
\end{equation}
and
\begin{equation}
G_0 =[ (E-E^i) - h_0 -h_i + i\varepsilon]^{-1}, \label{eq:2.18}
\end{equation}
is the free two-body propagator.
Finally, the two-body equation for the medium modified
two-body operator that appears in Eq.~(\ref{eq:2.12}) may be 
defined as
\begin{eqnarray}
{\cal T}_{0i}&\approx& {\cal T}_{2B} + {\cal T}_{2B} G_0 W_i G_i
{\cal T}_{0i} \nonumber \\
{\cal T}_{0i}&\approx& {\cal T}_{2B} + {\cal T}_{2B} G_0 {T}_{iC} G_0
{\cal T}_{0i},    \label{eq:2.19}
\end{eqnarray}
where
\begin{equation}
{T}_{iC}=W_i+W_i+G_0{T}_{iC},
\end{equation}
is the sum of all interactions between the excited nucleon
and the spectator core.

\begin{figure}
\includegraphics[width=8.5cm]{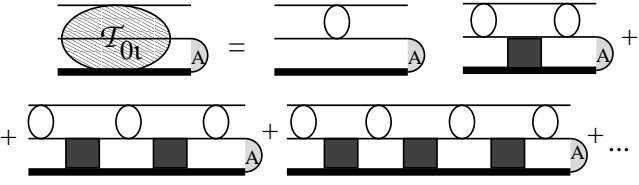}
\caption{A graphical depiction of Eq.~(\protect\ref{eq2.22}) which
represents the t-operator method for calculating the medium effects
for nucleon-nucleus scattering.
The nucleon-nucleon interaction
is represented by the ellipses (${\cal T}_{2B}$). The mean-field of the
core nucleus,
($T_{iC}$), is represented by the black squares. The spectator core
is represented by the
thick black line at the bottom which exists throughout the entire
calculation. The two top thin black lines represent the two active
nucleons which exist in the continuum of the nuclear well
(as depicted in Fig.~\protect\ref{fig4.1}). Only at the
beginning 
of the calculation does the ground state of the nucleus
exist as denoted by the semi-circle imprinted by an `A'.
There
is never an explicit interaction between the top-line projectile and
the spectator core in this theory.\label{fig4.2}}
\end{figure}
The projectile nucleon interacts with the
a struck target nucleon through ${\cal T}_{2B}$ and the nuclear
core interacts with this same stuck nucleon by
${T}_{iC}$. Therefore Equation~(\ref{eq:2.19}) can be written as a
set of two coupled equations
\begin{eqnarray}
{\cal T}_{0i}&=& {\cal T}_{2B} + {\cal T}_{2B} 
G_0 \hat{X_i}\nonumber \\
\hat{X_i}&=&T_{iC}G_0{\cal T}_{0i},\label{eq2.22a}
\end{eqnarray}
where the homogeneous equation represents the bound state and
the driving term in the first equation represents the projectile
interacting with the bound state.  Iterating Eq.~(\ref{eq:2.19})
we may understand
the physical mechanism better
\begin{eqnarray}
{\cal T}_{0i}|\Phi_A\rangle&=& {\cal T}_{2B}|\Phi_A\rangle
 \nonumber \\
&+& {\cal T}_{2B} G_0 {T}_{iC} G_0 {\cal T}_{2B}|\Phi_A\rangle
 \nonumber \\
&+& {\cal T}_{2B} G_0 {T}_{iC} G_0 {\cal T}_{2B}
G_0 {T}_{iC} G_0 {\cal T}_{2B}|\Phi_A\rangle  \nonumber \\
&+&\ldots, \label{eq2.22}
\end{eqnarray}
which is graphically represented in  Fig.~\ref{fig4.2}.
Equations(\ref{eq2.22a}-\ref{eq2.22}) is an
approximation to Faddeev's exact theory for 
three bodies~\cite{fadeev1} which contains three coupled equations.
It is approximate because the projectile never interacts with
the spectator core while in a true Faddeev three-body theory all three 
particles interact on equal footing with each other, this distinction
will be utilized later.
A good
summary of this approximate three-body theory (projectile, target
nucleon, and spectator core) can be found
in Refs.~\cite{bolle,med2,med1}. 

\subsection{Folding $g$ operator approach}\label{ssecg}
There is another popular approach to reducing the operator
${\cal G}_0{\cal Q}$ from a many-body operator to a two-body operator.
The two-body interaction, as stated previously, 
used in most nucleon-nucleus
interactions involves calculating Eq.~(\ref{eq4.9})
\begin{eqnarray*}
\hat{\tau_{0i}}=V_{0i}+V_{0i}{\cal G}_0Q\hat{\tau_{0i}}.
\end{eqnarray*}
In the traditional g-operator method (see for example
Refs.~\cite{amos1,hugoff,hugo1,hugo2})
this equation is simplified
to
\begin{equation}
\hat{\tau_{0i}}\approx g_{0i}=V_{0i}+V_{0i}
{\cal G_Q}g_{0i}, \label{eq4.20}
\end{equation}
where the propagator, ${\cal G}_0$ is modified to represent  the
medium of the nuclear bound state signified 
with the subscript ${\cal Q}$. The
$V_{0i}$ is still the bare nucleon-nucleon potential. As with the
t-operator approach it is usually assumed that there is only one
active target nucleon and the rest provide a
mean field~\cite{amos1}. The medium effects contained within the
propagator can be developed using various schemes. One method
is to treat the nucleus like an infinite Fermi gas and derive
spectral functions. Another is to include intermediate
Pauli-blocking
directly by using an operator which only allows entering
a state
which is above the Fermi level. This provides a similar
mechanism  to the folding $t$ approach
which includes only excited states in the
calculation via the projector ${\cal Q}$.  
The iterative form of the $g$ equation is
\begin{equation}
g_{0i}=V_{0i}+ V_{0i}{\cal G_Q}V_{0i}+
V_{0i}{\cal G_Q}V_{0i}{\cal G_Q}V_{0i}+ \ldots,
\label{eq4.22}
\end{equation}
which is similar in structure to Eq.~(\ref{eq2.22}) of the 
t-operator approach in which the total reaction can be broken
into interactions with either the two nucleons ($V_{0i}$) or 
the spectator core and the target nucleon ($\cal G_Q$) to first
order.
This theory
has a rich history which was begun by 
Brueckner~\cite{BBG1,BBG2} and is still 
active~\cite{amos1}. 

\begin{figure}
\includegraphics[width=8.5cm]{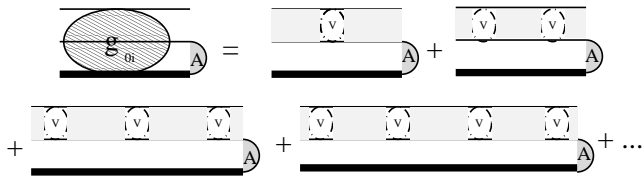}
\caption{A graphical depiction of Eq.~(\protect\ref{eq4.22}) which
represents the g-operator method of calculating the medium effects
for nucleon-nucleus scattering.
The nucleon-nucleon interaction
is represented by the ellipses ($V$). The propagator
has been modified to represent the mean-field effects of the
core nucleus (the grayed background).
The spectator core
is represented by the bottom
thick black line which exists throughout the entire
calculation. The two top thin black lines represent the two active
nucleons which exist in the continuum of the nuclear well
(as depicted in Fig.~\protect\ref{fig4.1}). Only at the
beginning (and also end for elastic scattering)
of the calculation does the ground state of the nucleus
exist as denoted by the semi-circle imprinted by an `A'.
There
is never an explicit interaction between the top-line projectile and
the spectator core in this theory.
\label{fig4.3}}
\end{figure}
In Fig.~\ref{fig4.3} a graphical depiction
of the g-operator theory is shown where
the free propagator has been `grayed' in to show that it has been
modified.
In g-operator theory the potential, $V_{0i}$, is the active operator
in contrast with t-operator theory which uses ${\cal T}_{2B}$ as  its
focus.
Both the $g$ operator method and the $t$ operator
method allow the modification of the two nucleon interaction
via changes in  the propagator. These changes
represent the effects of the nuclear medium that the
struck nucleon is bound to and the projectile is
moving through. These `medium effects' have been shown to
be important in a variety of different
calculations~\cite{amos1,med2,65mev}, notably below 200 MeV laboratory
energy for the incoming nucleon. In both formulations the 
interaction between the projectile and the spectator core is 
usually ignored. Again this approximation will be exploited in the
next section.

\section{Antisymmetrization}\label{sec3.3}

Watson and
collaborators~\cite {watson2,watson3}
concluded that the use of the antisymmetrized two nucleon interaction
was all that was required in high energy calculations for a correct
Pauli principle inclusion.
The argument was
qualitative in nature but powerful.
A chance of significant overlap of the projectile
wave function with more than one target nucleon is small for
scattering events with projectiles of high energy.
If overlap is inconsequential, the calculation may be reduced to
a two-body problem and thus
only two-body antisymmetrization is required~\cite{watson3}.
This will be referred to as the `Watson approximation' in the
remainder of this work.

\begin{figure}
\begin{center}
\includegraphics{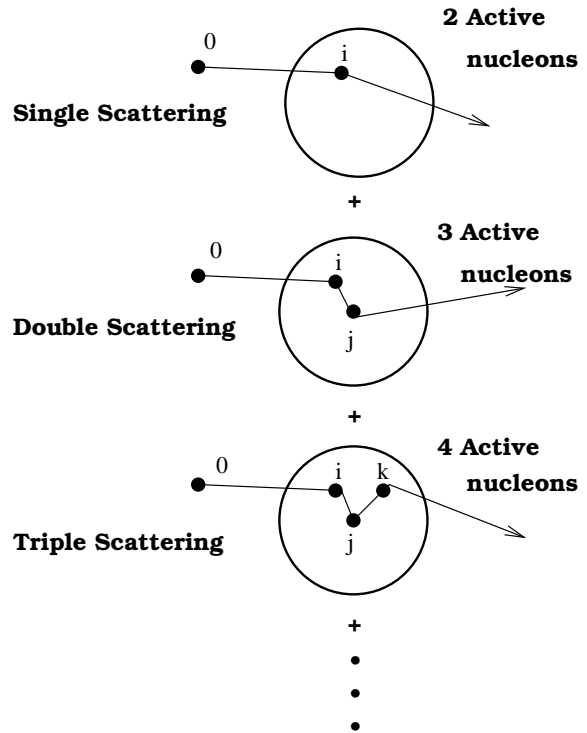}
\caption{An outline of the spectator expansion. The top
drawing represents the  first-order calculation with only
two active particles, the projectile and one target nucleon. The
second order calculation involves two target nucleons, etc.
\label{fig4.4}}
\end{center}
\end{figure}
Many modern microscopic scattering theories are organized using
a spectator expansion as depicted in Fig.~\ref{fig4.4}.
The first-order calculation only
scatters from nucleon $i$.  The second-order calculation
involves double-scattering events from nucleons $i$ and $j$.
The Watson approximation,
based on wave function overlap, is consistent with the
spectator expansion which only antisymmeterizes the active particles. 

Historically however there have been examples in which
a theory has been revised to include the spectators as part of the
antisymmetrization process.
For example
in nuclear physics 
the representation of the  ground
state of a nucleus improves when the wave function is 
completely antisymmetrized.
Most models only include interactions
of nearest neighbors, however
all  nucleons in the nucleus  interact
indirectly via the Pauli principle
and thus  the importance of an 
antisymmetrization scheme for the whole nucleus.

An older example is that of spin. Although
spin is added to the non-relativistic Schr\"{o}dinger
wave functions, the most significant role it plays in atomic
physics is to produce the correct fermionic
combinatorics by assigning each electron a unique
orbital. The actual
interactions which include spin are
relatively weak and are often ignored. 
It is with these precedents plausible to
develop a theory which includes both active and spectator
nucleons in its antisymmetrization scheme in contrast
to the Watson approximation. This is the 
goal of the next subsections where we will 
examine the role and identity
of the projectile in a many-body optical
potential scattering theory.

\subsection{A distinguishable projectile}\label{CM}
In traditional nucleon-nucleon scattering theory 
the label of projectile is often given to one of the nucleons
in the initial state.
This label  is not an intrinsic quality and therefore
not an adequate  quantum number. During
the scattering process the role of the projectile is not unique, its
identity is lost, and only by convention do we choose the
final state projectile to have the same isospin
projection as the initial projectile.

In exact Faddeev elastic nucleon-deuteron scattering the projectile
is also not an intrinsic label~\cite{deuteron}. 
The projectile
is defined, by convention,
in both the initial and final states as the nucleon
which is not bound.
During the intermediate states of 
the scattering process this  
characteristic is lost and all nucleons
are on equal identity footing in the Faddeev scheme. 

In nucleon-nucleus scattering 
is the projectile also indistinguishable?
A condition  for the projectile in nucleon-nucleus scattering
is that it is `fast'. This argument
was given by Watson and collaborators in Refs.~\cite{watson2,watson1}
and has some merit. On average, in the center of
mass frame of the nucleon-nucleus system, the free
projectile nucleon has a momentum magnitude
a factor of $A$ times greater than the target nucleon.
This constraint is however not sufficiently 
strong  to warrant quantum 
distinguishability because it is
only a mean kinematical expectation.

The strongest criteria for distinguishability of the
projectile comes from the
approximate form of the
theory itself. To create the optical potential
one sums up over all the target nucleons
(see Eq.~(\ref{eq4.11})). This summation could imply that the
target nucleon that is struck in the
optical potential theory  is actually the average composite 
nucleon
of the nucleus.
For example in an $N=Z$ nucleus the {\it average}
target nucleon is half proton and half neutron.
The theory, because of its approximate nature, has created an 
artificial  distinguishable composite nucleon target separable
from the projectile during the entire collision process.
Since this
composite
half-proton and half-neutron
target 
nucleon has no possibility to be considered identical with the
projectile then all traditional
kinematic exchange factors which double the strength of 
the transition operator for either proton-proton
or neutron-neutron scattering  
should be ignored as the composite  nucleus optical
potential is built via  Eq.~(\ref{eq4.11}).
The implication of this
is simply 
that all traditional neutron-neutron or 
proton-proton two-body amplitudes
are cut in half when used in summing the complete
many-body optical potential
\begin{eqnarray}
|\Psi_{CM-np}\rangle&&\equiv|\Psi_{2B-np}\rangle\nonumber\\
|\Psi_{CM-pp}\rangle&&\equiv\frac{|\Psi_{2B-pp}\rangle}{\sqrt{2}},
\end{eqnarray}
using the notation established in Eq.~(\ref{eq4.11a}).
This effective theory
is a manifestation of the
approximate averaging made in the optical potential theory. 
As an
example of a calculation using this modification we show the
proton-$^{16}$O optical potential given in traditional form by
Eq.~(\ref{eq4.14b}) is now modified to be
\begin{eqnarray}
&&{\cal P}U_{CM}{\cal P}\equiv\nonumber \\
&8&\int d{\bf {\tilde{k}'_i}}\; d{\bf {\tilde{k}_i}}\;
\langle \frac{\Psi_{2B-pp}}{\sqrt{2}} |{\tau}_{0i}
|\frac{\Psi_{2B-pp}}{\sqrt{2}}\rangle
\rho_{proton}({\bf {\tilde{k}'_i}},{\bf \tilde{k}_i}) \nonumber \\
+&8&\int d{\bf {\tilde{k}'_i}}\; d{\bf {\tilde{k}_i}}\;
\langle \Psi_{2B-np} |{\tau}_{0i}
|\Psi_{2B-np}\rangle
\rho_{neutron}({\bf {\tilde{k}'_i}},{\bf \tilde{k}_i}).\nonumber\\
\label{eq4.14d}
\end{eqnarray}
In
section \ref{sec3.6} we will show results of this type 
of calculation under the `CM' label (composite model).

This `composite model' theory does have disadvantages.
The theory should be able to be extended to
such processes as charge exchange reactions but it
no longer has the apparatus for their inclusion since each
target nucleon now only contains a fraction of charge. In the
next subsection we will look at a different, more appealing,
theoretical model for antisymmetrization.

\subsection{Projectile as quantum label}\label{MB}
In the first-order many-body
theory there is a difference
between the projectile nucleon and the struck target nucleon
which has not yet been exploited.
In Fig.~\ref{fig4.2}, the
$t$-folding operator model shows that the
struck target nucleon is acted upon by the mean field and thus
explicitly interacts with the nucleus while the
projectile does not. This differentiates the two nucleons
and is true throughout the whole
calculation.
The projectile nucleon at all
times carries the characteristic
that it is absent an interaction with the
spectator core. Conversely, the struck target nucleon 
interacts via a mean field with
the spectator core throughout the entire reaction. 
If the $g$-folding
theory has the projectile 
interact with the infinite nuclear matter of the core
than it contains higher order terms~\cite{amos2} and 
the projectile will be difficult to distinguish.
However if the
theory does not have the projectile
interact with the nuclear matter core,
as in Fig.~\ref{fig4.3},
than this same differential
characteristic
may be exploited.

Introducing a new quantum number formalism for the many-body
interaction basis,
where $|a,m_a\rangle$ is the addition, Eq.~(\ref{eq14}) now 
becomes
\begin{equation}
|\Psi_{MB}\rangle\equiv|\Psi_{2B}\rangle|a,m_a\rangle,
\label{eq5.1}
\end{equation}
where the $MB$ still represents a two-body interaction but taken within
the many-body context of nucleon-nucleus scattering.
The additional quantum number $a$
is a vector fundamentally belonging to the SU(2)
group for one nucleon (as do spin and isospin) which
denotes a new  intrinsic quality created by the
approximate form inherent in the optical potential theory:
$m_a=+\frac{1}{2}$ if an interaction
is not required with
the spectator core
and $m_a=-\frac{1}{2}$ if it is required.
In all
pure  first-order optical potential
elastic scattering theories the projectile has
$m_a=+\frac{1}{2}$ because there is no explicit interaction
with the spectator, 
all other target nucleons have
$m_a=-\frac{1}{2}$
because they interact via the mean field and
single particle density.

\begin{table}
\begin{center}
\begin{tabular}{||cc|cccc||}
\hline
$a$ & state & $m_a$ & pictorial & Coef.& example theory \\ \hline
 & & 1&$\uparrow\uparrow$ & 1 & nucleon-nucleon \\ \cline{3-6}
1  & symmetric & 0& $\uparrow\downarrow+\downarrow\uparrow$ & .5
& optical potential\\ \cline{3-6}
  & & -1& $\downarrow\downarrow$ & 1 & exact Faddeev
\\ \hline
0 & antisymmetric & 0& $\uparrow\downarrow-\downarrow\uparrow$ &.5
& optical potential \\ \hline
\end{tabular}
\caption{A description of the combination
of quantum number $a$ from the projectile and the struck
nucleon.  This quantum number represents an intrinsic quantity
of each nucleon involved in a many-body
scattering theory and belongs to the
SU(2) group (as does spin and isospin).
Coef. is the Clebsch-Gordan
coefficient squared and represents the probability. The
Faddeev~\protect\cite{faddeev} and
two-body nucleon-nucleon scattering theories are always symmetric
while the optical potential is mixed in symmetry  for
the $a$ quantum number
because of its approximate nature. }
\label{T4.1}
\end{center}
\end{table}
The $a$ quantum
number adds like the spin
and isospin quantum numbers, ${\bf a=a_0+a_i}$
and $m_a={m_a}_0+{m_a}_i$. An analogy to isospin can be clearly
elucidated. The most significant difference between isospin
flavors (neutron or proton) 
is the effect of the  coulomb interaction, likewise the difference
between $a$ flavors (projectile and non-projectile) is the 
effect of the spectator core nucleon's interaction. 
In Table~\ref{T4.1} we summarize the
combinatorics of this new vector.
Importantly
this new many-body basis does not change the nature of the
fundamental
nucleon-nucleon interaction,
it is only a placeholder vector symbolizing 
the  status of the interaction
with the spectator core. The reason for its existence is solely
because the Faddeev theory is approximated in developing 
the optical potential, the additional quantum numbers are the relics
of that approximation.
The Hamiltonian for the
nucleon-nucleon interaction conserves this new quantum number and
the measurement of it commutes with all dynamical variables.  The
significance of this quantum number is the distinguishability
it produces is needed in the many-body context.
This becomes clearer if we re-examine the t-operator method of
Eq.~(\ref{eq2.22}):
\begin{eqnarray}
{\cal T}_{0i}|\Phi_A\rangle&=& {\cal T}_{2B}|\Phi_A\rangle
 \nonumber \\
&+& {\cal T}_{2B} G_0 {T}_{iC} G_0 {\cal T}_{2B}|\Phi_A\rangle
 \nonumber \\
&+& {\cal T}_{2B} G_0 {T}_{iC} G_0 {\cal T}_{2B}
G_0 {T}_{iC} G_0 {\cal T}_{2B}|\Phi_A\rangle  \nonumber \\
&+&\ldots \nonumber
\end{eqnarray}
The operator ${\cal T}_{2B}$ represents the fundamental nucleon-nucleon
interaction and although a quantum number has been added this interaction
remains unchanged
\begin{eqnarray}
&&\sum_T\langle a=1,m_a=0|\langle\Psi_{2B}|{\cal T}_{2B}
|\Psi_{2B}\rangle|a=1,m_a=0\rangle\nonumber\\
&\equiv& 
\sum_T\langle a=0,m_a=0|\langle\Psi_{2B}|{\cal T}_{2B}
|\Psi_{2B}\rangle|a=0,m_a=0\rangle\nonumber\\
&\equiv&
.5\;\sum_T\langle\Psi_{2B}|{\cal T}_{2B}
|\Psi_{2B}\rangle,
\end{eqnarray}
assuming that the nucleon-nucleon potential is isospin ($T$)
independent. The factor of .5 recognizes that the new quantum
number, $a$, divides the original space in two, 
a symmetric ($a=1$) and antisymmetric ($a=0$) part.

In every final channel of the ${\cal T}_{2B}$ operator the
struck target
nucleon must be differentiated from the projectile  to calculate
the ${T}_{iC}$ operator, which is the interaction between the
target nucleon and the spectator core
\begin{eqnarray}
\biggl\langle a_i=\frac{1}{2},{m_a}_i=+\frac{1}{2}\biggm|T_{iC}\biggm|
a_i=\frac{1}{2},{m_a}_i=+\frac{1}{2}\biggr\rangle= 0\nonumber\\
\biggl\langle a_i=\frac{1}{2},{m_a}_i=-\frac{1}{2}\biggm|T_{iC}\biggm|
a_i=\frac{1}{2},{m_a}_i=-\frac{1}{2}\biggr\rangle\neq 0, \nonumber \\
\end{eqnarray}
where the first equation represents the interaction,
with null result,
of the projectile with the core $A-1$ nucleons while
the second equation represents the target nucleon interacting
with the core.
Likewise in the initial and final elastic channels of the complete
reaction the struck nucleon is bound using the $|\Phi_A\rangle$
basis to describe the nucleus. In this modified version we 
add the $a_i,{m_a}_i$ quantum number description to this basis in which
only $m_a=-\frac{1}{2}$ is non zero.

This new two-body $|a,m_a\rangle$ eigenvector
is always symmetric for nucleon-nucleon scattering $(a=1,m_a=-1)$
where there are no other interaction requirements 
and it is also always symmetric for
Faddeev nucleon-deuteron scattering ($a=1,m_a=+1$)
where all interaction requirements are explicitly included.
Thus because both nucleons have the same projections in these
exact theories all protons
can still be considered identical.
Conversely, all approximate first-order optical
potential theories
allow the $|a,m_a\rangle$ eigenvector
to be mixed symmetric (50\%) or antisymmetric (50\%) with an
$m_a=0$. This is an additional degree of freedom introduced
specifically by the approximation in the first-order
optical potential theory.

Because
the optical potential theory distinguishes between projectile
and target nucleons the new quantum $a$ number may be
exchanged either
symmetrically or antisymmetrically as long as the overall
scattering amplitude
remains antisymmetric. 
In the Watson approximation
the states represented by $a=1$ are the only physical states
included. The many-body theory, presented in this work, allows
both $a=0$ and $a=1$ states.
Its behavior in the many-body optical potential theory 
fundamentally changes the acceptable basis from which it operates.

Applying this new quantum number to neutron-proton scattering
in the many-body context
it is again noted that the addition does not change the strength of
the two nucleon scattering phase space 
because this quantum
number represents an external interaction. 
The new neutron-proton scattering amplitude in a many-body
context now has a total of eight
states: 
\begin{eqnarray}
&&|{\bf 1_{MB}}\rangle\equiv
|{\bf q},{\bf K}\rangle_{sym}|S=0,m_s\rangle
|T=1,0\rangle|a=1,0\rangle\nonumber\\
&&|{\bf 2_{MB}}\rangle\equiv
|{\bf q},{\bf K}\rangle_{asym}|S=1,m_s\rangle
|T=1,0\rangle|a=1,0\rangle\nonumber \\
&&|{\bf 3_{MB}}\rangle\equiv
|{\bf q},{\bf K}\rangle_{sym}|S=1,m_s\rangle
|T=0,0\rangle|a=1,0\rangle\nonumber \\
&&|{\bf 4_{MB}}\rangle\equiv
|{\bf q},{\bf K}\rangle_{asym}|S=0,m_s\rangle
|T=0,0\rangle|a=1,0\rangle\nonumber\\
&&|{\bf 5_{MB}}\rangle\equiv
|{\bf q},{\bf K}\rangle_{sym}|S=1,m_s\rangle
|T=1,0\rangle|a=0,0\rangle\nonumber\\
&&|{\bf 6_{MB}}\rangle\equiv
|{\bf q},{\bf K}\rangle_{asym}|S=0,m_s\rangle
|T=1,0\rangle|a=0,0\rangle\nonumber \\
&&|{\bf 7_{MB}}\rangle\equiv
|{\bf q},{\bf K}\rangle_{sym}|S=0,m_s\rangle
|T=0,0\rangle|a=0,0\rangle\nonumber \\
&&|{\bf 8_{MB}}\rangle\equiv
|{\bf q},{\bf K}\rangle_{asym}|S=1,m_s\rangle
|T=0,0\rangle|a=0,0\rangle\nonumber\\
\label{list2}
\end{eqnarray}
which
are antisymmetric. The first number in the
isospin and antisymmetric bra-kets represent
the full vector, the second number represents the
projection. All eight states 
are individually half the strength of
the four for two-body scattering listed in Eq.~(\ref{list1})
because the new quantum number $a$ splits the 
traditional space into two and renormalizes 
them (the coefficient of .5 in Table~{\ref{T4.1}}).
If the nucleon-nucleon potential
is isospin independent
(if the $T=1$ state
amplitudes are the same  as the $T=0$ amplitudes given an identical
momentum-spin space)
then the eight states reduce to only four
unique states.
The combined character and strength of the
four states are then the same for the traditional two-body
and many-body neutron-proton interactions.
The scattering amplitude therefore does not change with the
addition of the new quantum number in the neutron-proton case,
it is only an additional placeholder. The
proton and neutron have already been differentiated by isospin so this
new addition is redundant and inconsequential as expected.

The many-body proton-proton
scattering amplitude has a total of four states which
are antisymmetric (those that have T=1 designation only) 
\begin{eqnarray}
&&|{\bf 1_{MB}}\rangle\equiv
|{\bf q},{\bf K}\rangle_{sym}|S=0,m_s\rangle
|T=1,1\rangle|a=1,0\rangle\nonumber\\
&&|{\bf 2_{MB}}\rangle\equiv
|{\bf q},{\bf K}\rangle_{asym}|S=1,m_s\rangle
|T=1,1\rangle|a=1,0\rangle\nonumber \\
&&|{\bf 5_{MB}}\rangle\equiv
|{\bf q},{\bf K}\rangle_{sym}|S=1,m_s\rangle
|T=1,1\rangle|a=0,0\rangle\nonumber\\
&&|{\bf 6_{MB}}\rangle\equiv
|{\bf q},{\bf K}\rangle_{asym}|S=0,m_s\rangle
|T=1,1\rangle|a=0,0\rangle\nonumber\\
\label{list3}
\end{eqnarray}
which for comparison use the same numbering scheme as
the neutron-proton states listed in Eq.~(\ref{list2}).
These states have distinct differences from their
pure nucleon-nucleon counterparts. Two of the states
include even momentum space-spin space 
product wave functions which were not
allowed in the traditional two-body proton-proton case.
States $|{\bf 5_{MB}}\rangle$ and $|{\bf 6_{MB}}\rangle$ include
formally forbidden scattering states like 
$^3S_1$ and $^1P_1$ written in traditional
partial wave $^{2S+1}L_J$ notation.
Also there is another significant change for proton-proton
scattering in the many-body context.
The protons are no longer considered identical thus the
traditional kinematical doubling for identical
particles described earlier will no longer occur.
The protons are no longer
identical because of the new $a$ quantum number,
thus they are now on the same footing as the neutron-proton interaction 
kinematically.

In summary this new quantum number, to first approximation,
does not change the character of the neutron-proton interaction
but the proton-proton interaction now mimics the neutron-proton
interaction at half strength (only half the number of states are possible)
\begin{eqnarray}
|\Psi_{MB-np}\rangle&&\equiv|\Psi_{2B-np}\rangle\nonumber\\
|\Psi_{MB-pp}\rangle&&\equiv\frac{|\Psi_{2B-np}\rangle}{\sqrt{2}}.
\end{eqnarray}
This is a manifestation
of the approximate nature of the first order optical
potential and is not a characteristic of the fundamental force. As an
example of a calculation using this modification we show the 
proton-$^{16}$O optical potential given in traditional form by
Eq.~(\ref{eq4.14b}) is now modified to be
\begin{eqnarray}
&&{\cal P}U_{MB}{\cal P}\equiv\nonumber \\
&8&\int d{\bf {\tilde{k}'_i}}\; d{\bf {\tilde{k}_i}}\;
\langle \frac{\Psi_{2B-np}}{\sqrt{2}} |{\tau}_{0i}
|\frac{\Psi_{2B-np}}{\sqrt{2}}\rangle
\rho_{proton}({\bf {\tilde{k}'_i}},{\bf \tilde{k}_i}) \nonumber \\
+&8&\int d{\bf {\tilde{k}'_i}}\; d{\bf {\tilde{k}_i}}\;
\langle \Psi_{2B-np} |{\tau}_{0i}
|\Psi_{2B-np}\rangle
\rho_{neutron}({\bf {\tilde{k}'_i}},{\bf \tilde{k}_i}).\nonumber\\
\label{eq4.14c}
\end{eqnarray}
The results of nucleon-nucleon scattering and exact Faddeev scattering
are left unaffected by this work as expected.

Exchange of ${m_a}$ is not
included explicitly in the many-body amplitude but it is assumed
to exist.
Since the $m_a$ quantum number is tied directly
with the identity of the projectile in 
proton-proton (or neutron-neutron)
scattering 
it shares
many similarities with the $m_t$ quantum number in neutron-proton
scattering. In two-body scattering the 
significance of isospin exchange does not affect the final result. The
distinguishable calculation (not including isospin) is the exact
same as the indistinguishable result (including isospin) because
the projectile nucleon by convention 
always carried the same isospin component.
The same is true with this new quantum number $a$
in the many-body case. The projectile always
carries the $m_a=+\frac{1}{2}$
component so target exchange is irrelevant to
the final result. Irrelevant does not mean that
it does not happen, 
there is no need to add an explicit
exchange mechanism for quantum label $m_a$
to the nucleon-nucleus theory. This 
feature of exchange was noted
by Kowalski in Ref.~\cite{kowalski-rev}.

There have been two new antisymmetrization models presented. The
first, the composite model (CM), treats the target as 
distinguishable and
therefore not identical.
The second model  adds a new quantum
number $a$ to treat two protons
as indistinguishable but no longer identical.
The latter model is more sophisticated,
involves exchange in a natural way,
and treats antisymmetrization with a many-body formalism (MB). 
All three
models have the same neutron-proton amplitudes
but are differentiated by how they treat the proton-proton 
(and neutron-neutron) amplitudes.
In the next section we will show comparisons between the traditional
Watson approximation,
the composite model,
and the many-body
model.

\section{Results}\label{sec3.6}
Comparisons will be made between experimental data
and calculations which involve a folding, energy fixed,
$t$ operator approach
(described in section~\ref{ssect} and further in Ref.~\cite{density}).
The medium effects ($T_{iC}$) have 
been set to zero so that only the Born term
is calculated in Fig.~\ref{fig4.2}. The reasoning for not including
any medium modifications
is so that the differences between the strengths of the
antisymmetric techniques
can be clearly elucidated. Medium affects
are only significant with projectile energies
below 200 MeV~\cite{med1} so most comparisons
in this section are made
with energies of at least 200 MeV. To further ascertain the
significance of the antisymmetric technique
the same  Dirac-Hartree nuclear structure calculation~\cite{DH2} and
the same nucleon-nucleon potential~\cite{nijmegen}, 
fit to the 1993 dataset of 
nucleon-nucleon observables, is
used in all calculations unless otherwise noted.
Beyond these two inputs
there are no adjustable parameters.

\begin{figure*}
\includegraphics[width=10.75cm]{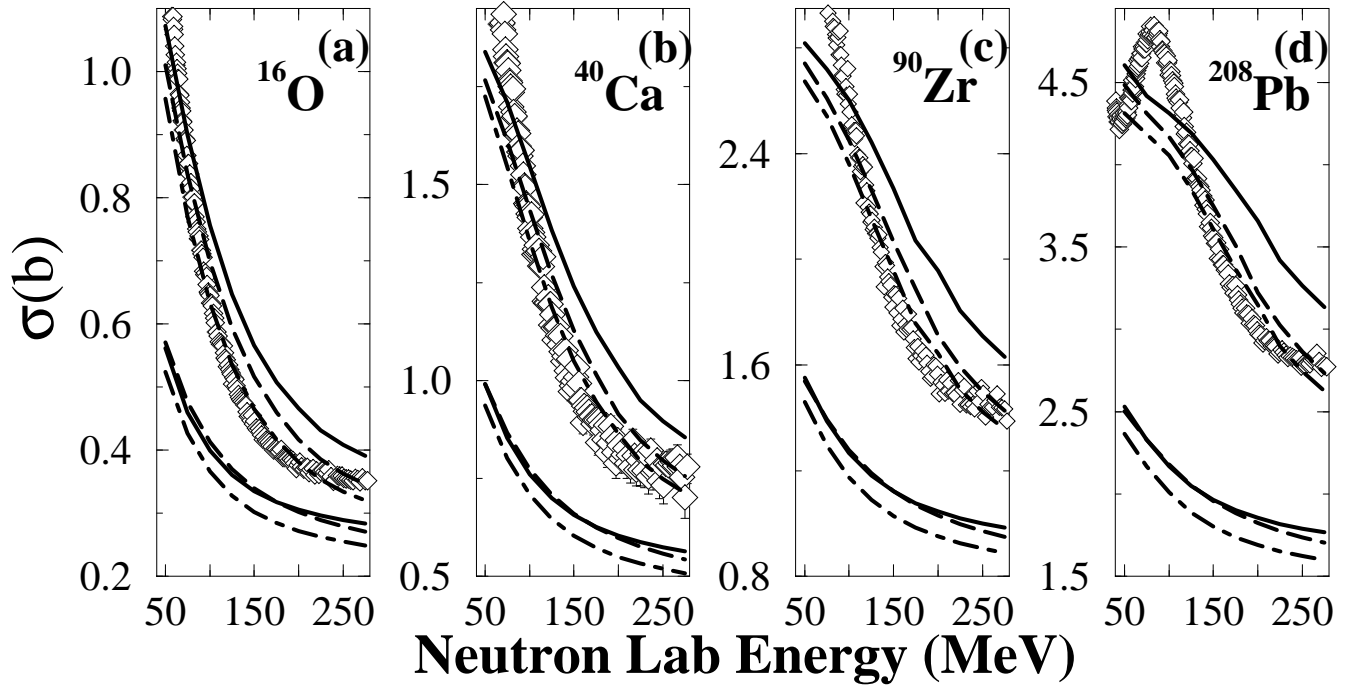}
\caption{Four total and inelastic (reaction) neutron-nucleus
cross section from four  different nuclear targets.
The upper lines are
calculations of the total cross section while the lower lines
represent the inelastic cross section.
The solid
line is the two-body Watson approximation,2B,
the dashed line is the
many-body calculation, MB,
and the dashed-dotted line
represents
the composite model (CM).
All calculations
use a Dirac-Hartree density~\protect\cite{DH2} and the
Nijmegen nucleon-nucleon potential~\protect\cite{nijmegen} and
a full-folding procedure~\protect\cite{density}.
The neutron total cross section
data (diamonds) are taken from Ref.~\protect\cite{finlay1}. There
is systematic error which is roughly represented by the size of
the diamonds~\protect\cite{finlay2}.
\label{fig3.1}}
\end{figure*}
The Watson approximation will be referred to as `2B' 
(discussed first in section~\ref{2B}), 
the `CM' will be the composite model
(discussed in section~\ref{CM}),
and the many-body
antisymmetry will be `MB'(discussed in section~\ref{MB}). 
To summarize the differences between the three theoretical models, 
the 2B model uses the same two-body interaction
used in nucleon-nucleon and nucleon-deuteron scattering, the CM model
cuts all proton-proton and neutron-neutron
amplitudes in half, and the MB model replaces
the proton-proton and neutron-neutron 
amplitudes with the neutron-proton amplitudes and
also cuts them in half. The end result  is that the strength of the
$m_t=\pm1$ amplitudes developed in this work are roughly half that
of the the traditional Watson two-body amplitudes. It will
be shown that this dramatic reduction leads to an improvement
in the theoretical fit to experimental data.  Then an
examination of why the Watson approximation has been successful 
for five decades and why now the need for modification.

In Fig.~\ref{fig3.1} we compare these three calculations
(2B,MB,CM)
on neutron-nucleus total and inelastic
cross sections  for four doubly magic
targets, $^{16}$O,$^{40}$Ca,$^{90}$Zr, and $^{208}$Pb.
The over prediction of the
Watson approximation (solid line)
in the range of validity for this
calculation ($\ge$ 200 MeV) stands out while the
other calculations (MB:dashed, CM:dashed-dotted)
do much better.
This over prediction for the total cross section
using the traditional Watson approximation
signifies that
the strength of the effective interaction designated
by the optical potential is much too large.
The data and calculation have a high
level of precision such that this difference is indeed real
and significant (sometimes as high as 20\%). The many-body and
composite model do much better by reducing the strength of the
neutron-neutron amplitudes which are inherent in the 
nucleon-nucleus optical
potential calculation.  

The inelastic cross section calculations are
included for future reference. An interesting facet
worthy of study is that surprisingly the two-body calculation
roughly matches the many-body result (but not the composite
result). Therefore the many-body result mainly reduces 
only the
elastic cross section while the composite model reduces
amplitudes which contribute to 
both the elastic and inelastic cross section.

\begin{figure}
\includegraphics[width=8.5cm]{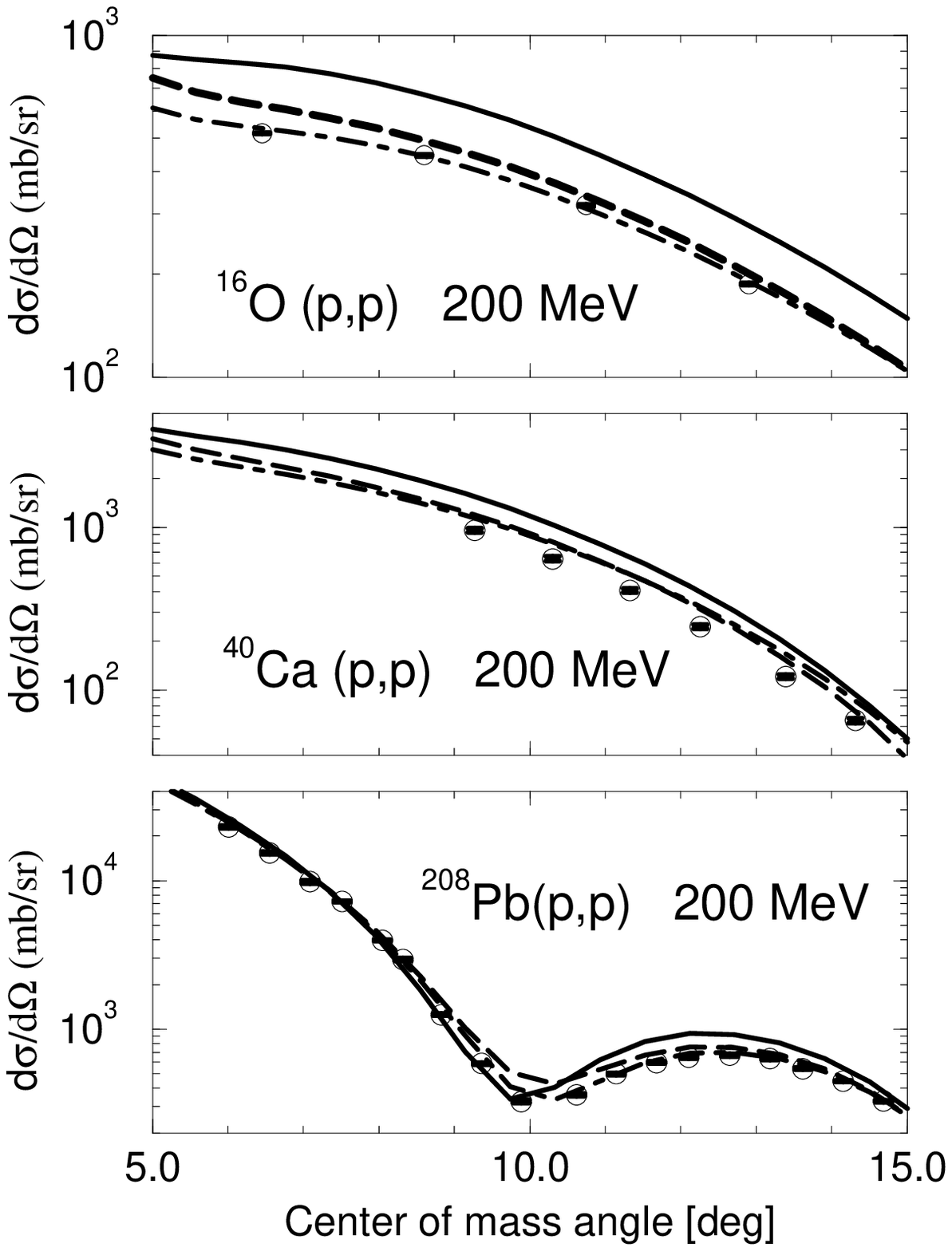}
\caption{
Extreme forward differential cross section calculations
and experimental data for a proton on
$^{16}$O,
a proton on
$^{40}$Ca,
and
a proton on
$^{208}$Pb
at 200 MeV proton laboratory energy.
The calculation procedure and inputs
are the same as Fig.~\protect\ref{fig3.1}.
The solid line represents the Watson
approximation (2B)
while the dashed line represents
the many-body antisymmetrization (MB)
and the dashed-dotted line represents the
composite model (CM).
The data points are
represented by circles and
are taken from Refs.~\protect\cite{ca200,pb200,qdata}.
\label{fig3.2}}
\end{figure}
Proton-nucleus differential cross section calculations
at the extreme forward angles for
$^{16}$O,$^{40}$Ca, and $^{208}$Pb
at a proton lab energy of 200 MeV are shown
in Fig.~\ref{fig3.2}.
Again the 2B Watson calculation (solid line) over-predicts
the data points at the extreme forward angles while
the other calculations (MB:dashed, CM:dashed-dotted)
compare better with experiment.
Although these
differences may appear slight they are actually rather significant
because this is a  logarithmic graph. Once again the use of the Watson
approximation (2B) leads to an
effective nucleon-nucleon
potential which has too large a strength (as much as 50\%!).
Both the many-body antisymmetrization (MB)
and the composite model (CM) come closer to the experimental values
in all cases tested at and above
this energy. Because these are proton projectiles
a coulomb force has been added~\cite{coulomb}. The systematic
use of first neutrons (in Fig.~\ref{fig3.1}) and then
protons (in Fig.~\ref{fig3.2}) as projectiles, resulting in the same
over prediction for the Watson approximation calculation,
removes the coulomb interaction as a source of the discrepancy.

\begin{figure}
\includegraphics[width=8.5cm]{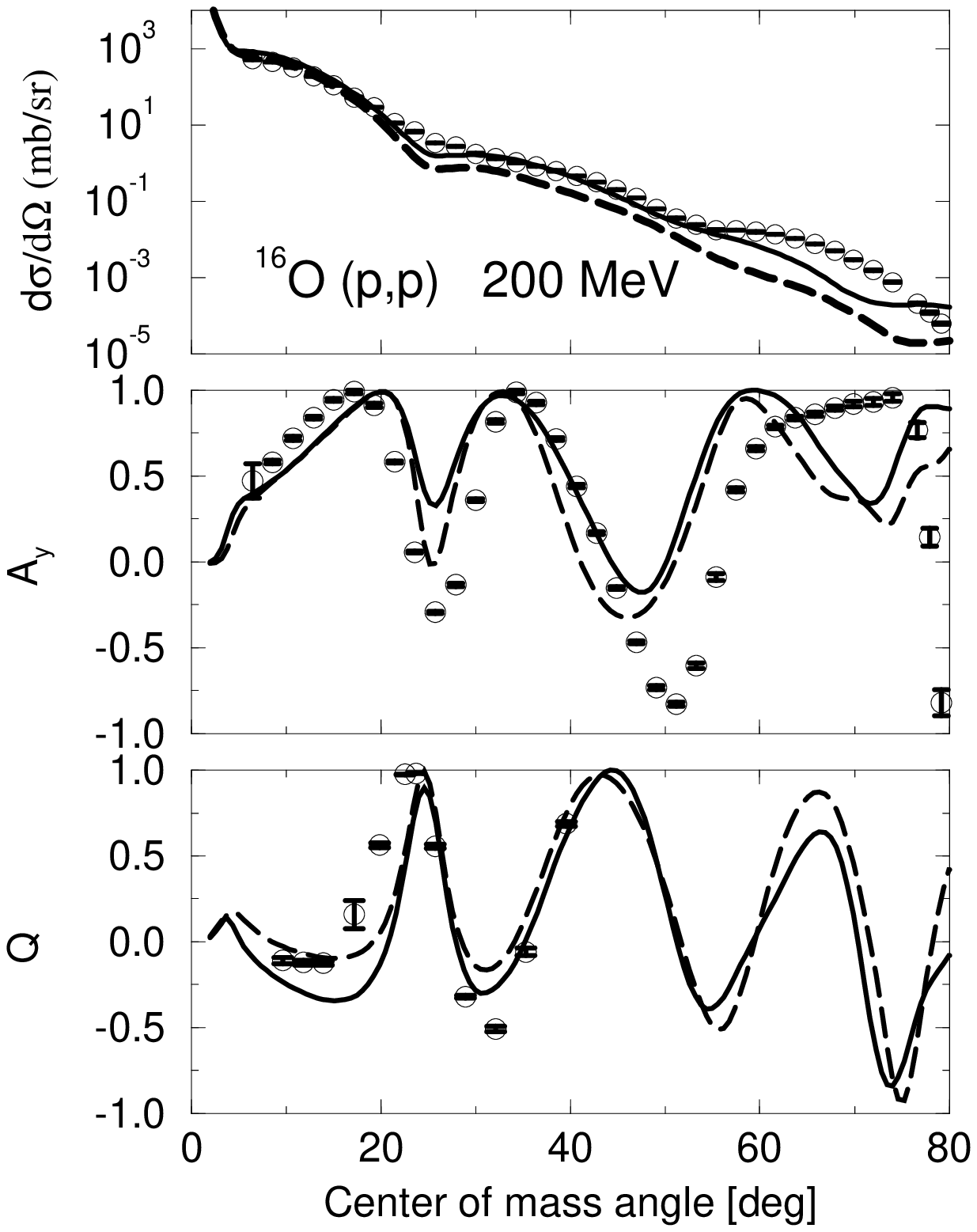}
\caption{ The differential cross section ($d\sigma/d\Omega$),
analyzing 
power or polarization ($A_y$), and the spin transfer ($Q$)
for a proton interacting with 
$^{16}$O at a laboratory energy of 200 MeV.
The data are the circles and were taken from Ref.~\cite{ca200}.
The solid line represents the Watson approximation (2B) while the
dashed line is the many-body antisymmetrization model (MB). For
graphical clarity the composite model (CM), which closely mimics MB,
is not included. The calculation methods and inputs are described in
Fig.~\protect\ref{fig3.1} and the text.
\label{fig3.5}}
\end{figure}

\begin{figure}
\includegraphics[width=8.5cm]{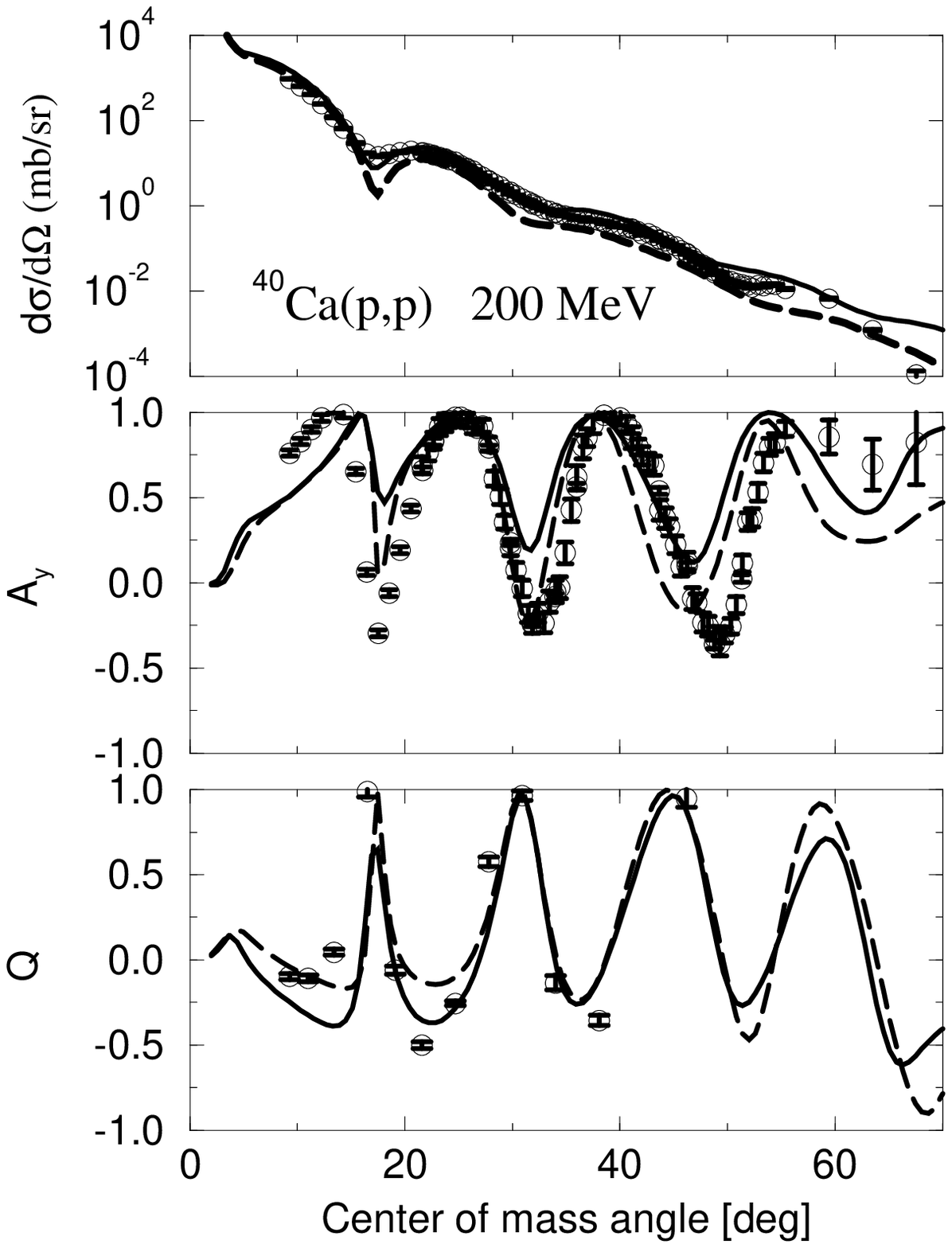}
\caption{
The differential cross section ($d\sigma/d\Omega$),
analyzing power or polarization ($A_y$), and the spin transfer ($Q$)
for a proton interacting 
with $^{40}$Ca at a laboratory energy of 200 MeV.
The data are the circles and were taken from Ref.~\cite{ca200}.
The solid line represents the Watson approximation (2B) while the
dashed line is the many-body antisymmetrization model (MB). For
graphical clarity the composite model (CM), which closely mimics MB,
is not included. The calculation methods and inputs are described in
Fig.~\protect\ref{fig3.1} and the text.
\label{fig3.6}}
\end{figure}

\begin{figure}
\includegraphics[width=8.5cm]{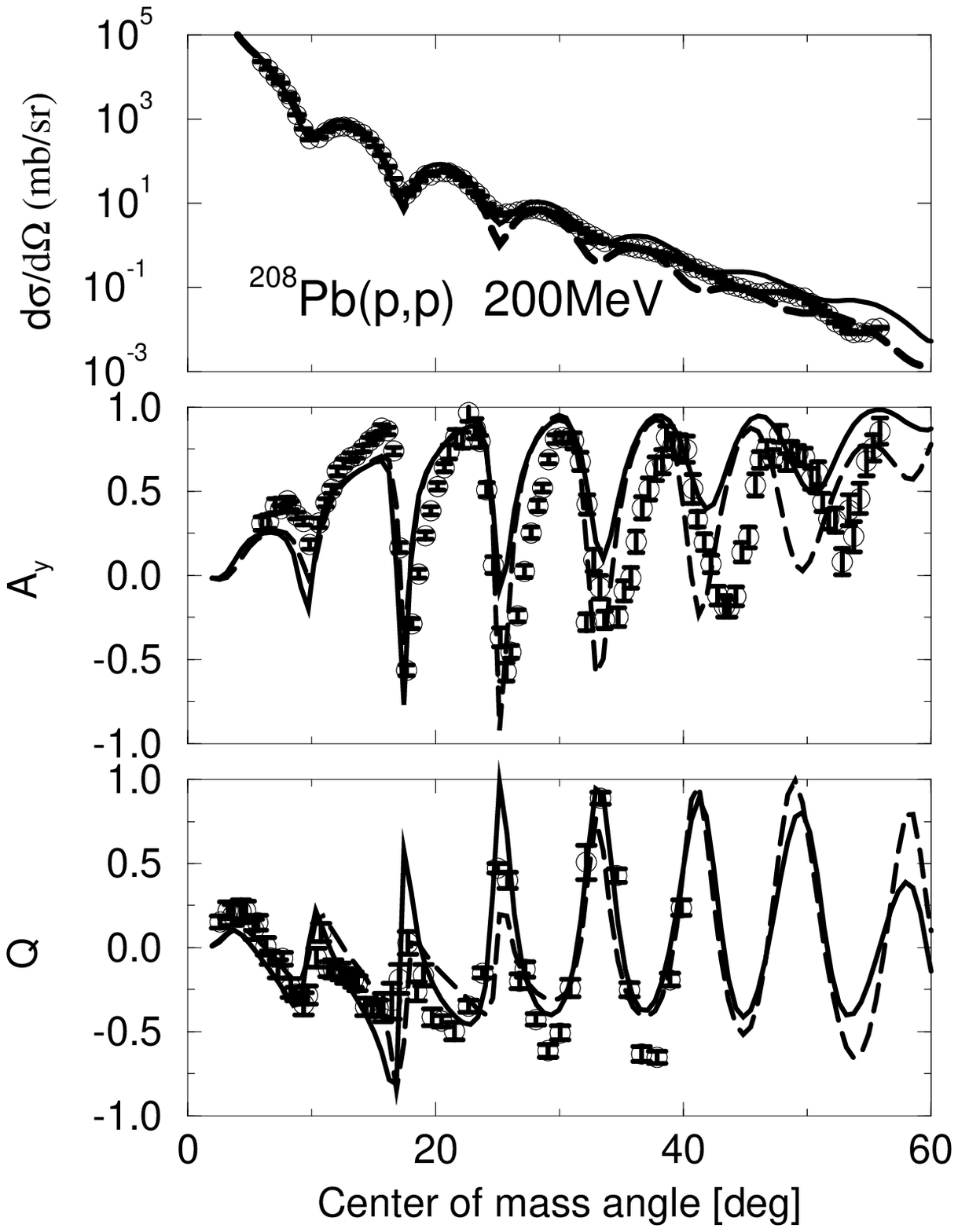}
\caption{
The differential cross section ($d\sigma/d\Omega$),
analyzing power or polarization ($A_y$), and the spin transfer ($Q$)
for a proton interacting with $^{208}$Pb at a laboratory
energy of 200 MeV.
The data are the circles and were taken from Ref.~\cite{ca200}.
The solid line represents the Watson approximation (2B) while the
dashed line is the many-body antisymmetrization model (MB). For
graphical clarity the composite model (CM), which closely mimics MB,
is not included. The calculation methods and inputs are described in
Fig.~\protect\ref{fig3.1} and the text.
\label{fig3.7}}
\end{figure}

The importance of fitting accurately the forward angles for 
proton-nucleus elastic differential cross sections must be 
emphasized. It is the most significant part of the cross section and 
therefore of importance to understand if the strength is to be 
calculated accurately using an effective interaction. 
Too often these discrepancies
are missed because of the range and logarithmic nature of the 
experimental measurements. For example these differences are barely
noticeable when the same data and calculations are graphed over a larger
range as in Figs.~\ref{fig3.5}-\ref{fig3.7}. 

The complete elastic experimental observables
for a proton scattering off of
$^{16}$O,$^{40}$Ca, and $^{208}$Pb respectively are shown 
in Figs.~\ref{fig3.5}-\ref{fig3.7}. The
top graph in each figure is the same differential cross section
that was depicted in Fig.~\ref{fig3.2} except now a
larger angle range is used. The middle graph is the spin
polarization ($A_y$) and
the bottom graph is the spin transfer ($Q$).  These bottom two graphs
are spin observables and measure the frequency of spin changes along
the axis of quantization and orthogonal to it respectively. They are
normalized to the elastic differential cross section.

The solid line is the
Watson approximation (2B), the dashed line is the many-body antisymmetry
technique (MB). The composite model (CM) is not 
shown for graphical clarity
but it mimics very closely the results of MB. The
new calculations described in this  work (MB and CM) do
as good a job as the Watson approximation (2B)
in the describing the full differential cross section and
spin observable experimental data. 
This result is somewhat surprising because the
modification to the two-body interaction was substantial. In both
of the new theories the strength of the proton-proton 
and neutron-neutron interaction was cut
in half but the spin observables, which are a ratio, seem to be 
relatively insensitive to this
modification keeping their same general oscillatory form. 

Overall the newly introduced composite model and many-body
antisymmetrization calculation predict the strength of the two-body
nuclear interaction used in a microscopic first-order
nucleon-nucleus elastic optical
potential better than the older Watson approximation. This
is over four different nuclei targets, two different projectiles,
and four different observables.

The Watson approximation has been accepted since the 1950's. In the late
1970's and early 1980's there were attempts at improving the Watson
approximation but ultimately they failed, in part because the
approximation did a very good job at reproducing experimental
results~\cite{ray1}. This paper suggests two
alternatives which  now reproduce experimental data better and are
more in the spirit of a many-body nucleon-nucleus theory. In
this section the success of the Watson approximation will be examined
and it will be shown that its poor quality has not been obvious until
recently.

The quality of nucleon-nucleus elastic scattering data
has changed in the past twenty years and this is due directly to
the dynamic nature of the nucleon-nucleon dataset.
In the 1980's the Bonn nucleon-nucleon potential~\cite{bonn}
was finalized. At the time of
publication it was fit to a 1986 dataset of nucleon-nucleon
observables. This refinement
continued in the early 1990's with the creation
of the Nijmegen potential~\cite{nijmegen}
and a newer Bonn potential~\cite{cdbonn}. There were
theoretical developments that these new potentials included, but more
importantly the world dataset of nucleon-nucleon observables
expanded and was modified~\cite{nndata}. In early 2000
the Bonn potential was once again modified
including the use of
an even
larger nucleon-nucleon dataset that included experiments
run as late as 1999~\cite{cdbonn2000}. The differences between the
1986 dataset and the 1993 dataset are quite dramatic and this abrupt
change led to a significant difference in results of nucleon-nucleus
elastic scattering calculations.

\begin{figure}
\includegraphics[width=8.5cm]{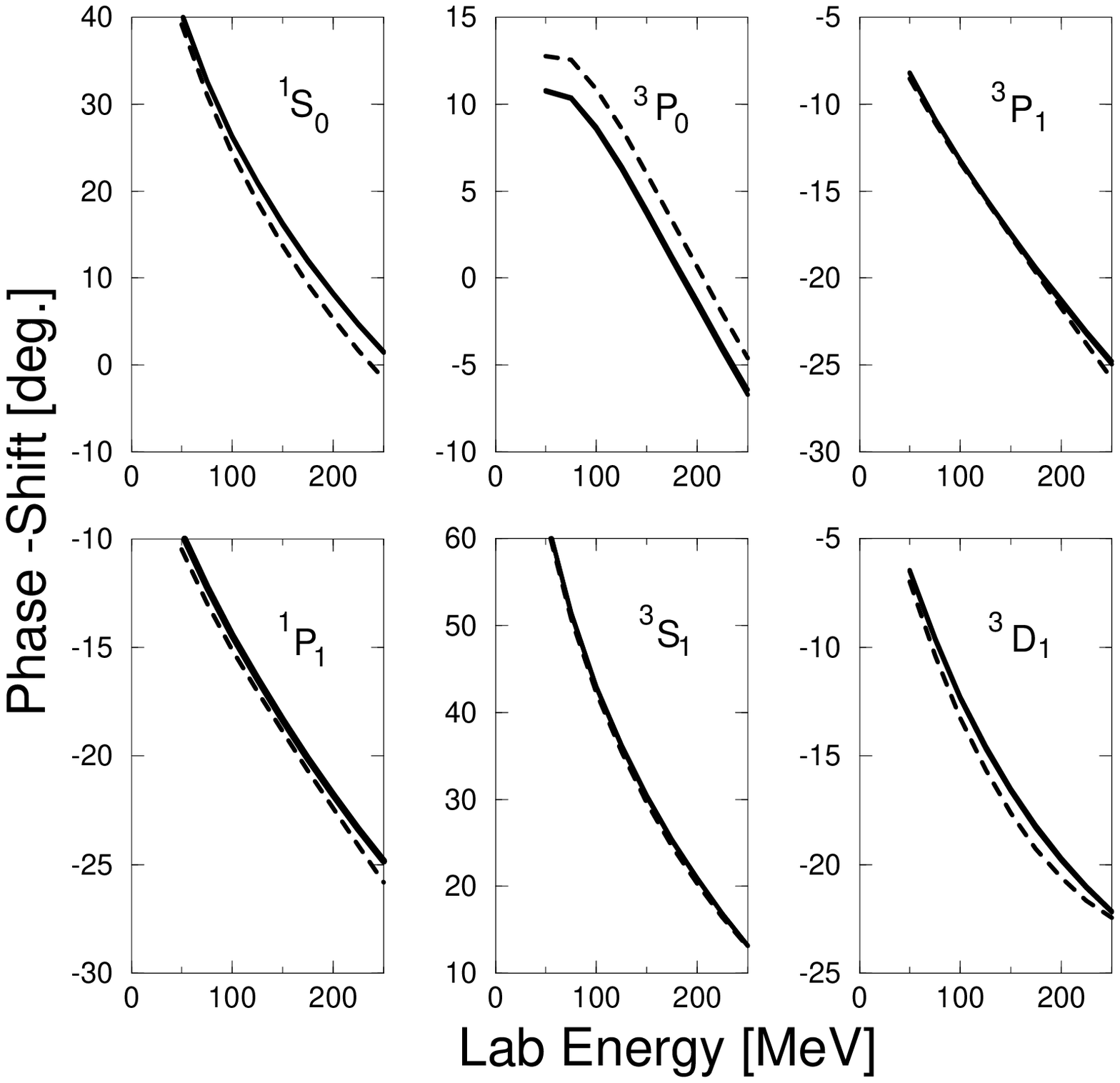}
\caption{The first six nucleon-nucleon phase-shifts as a function
of energy. The solid line represents a fit to the 1999
dataset~\protect\cite{cdbonn2000} while the dashed line represents a
fit to the 1986 dataset~\protect\cite{bonn}.
\label{fig3.8}}
\end{figure}
To explicitly see these differences 
the first few phase shifts are plotted
for the neutron-proton amplitude in Fig.~\ref{fig3.8}.
The solid line represents the
phase shift which fits the 1999 dataset. The dashed line represents
the phase shifts for the dataset thirteen years earlier in 1986.
Specifically for the $1S_0$ phase-shift these differences are quite
severe. The strength of the neutron-proton amplitude has changed
dramatically in this time period accounting for the differences
in calculation results for the two nucleon, three nucleon and many
nucleon scattering problems.  These changes over time have
led to a better description of nucleon-nucleon scattering,
nucleon-deuteron scattering (see for example Ref.~\cite{deuteron}), 
but a worsening of the nucleon-nucleus
scattering description
using a microscopic optical potential with the Watson
approximation. This is in concurrence with the focus of this work; that
the nucleon-nucleon amplitude is used correctly in 
nucleon-nucleon scattering, and in an exact
Faddeev scheme like nucleon-deuteron scattering~\cite{deuteron}, 
however 
it needs 
modification in the approximate optical potential theory.

\begin{figure}
\includegraphics[width=8.5cm]{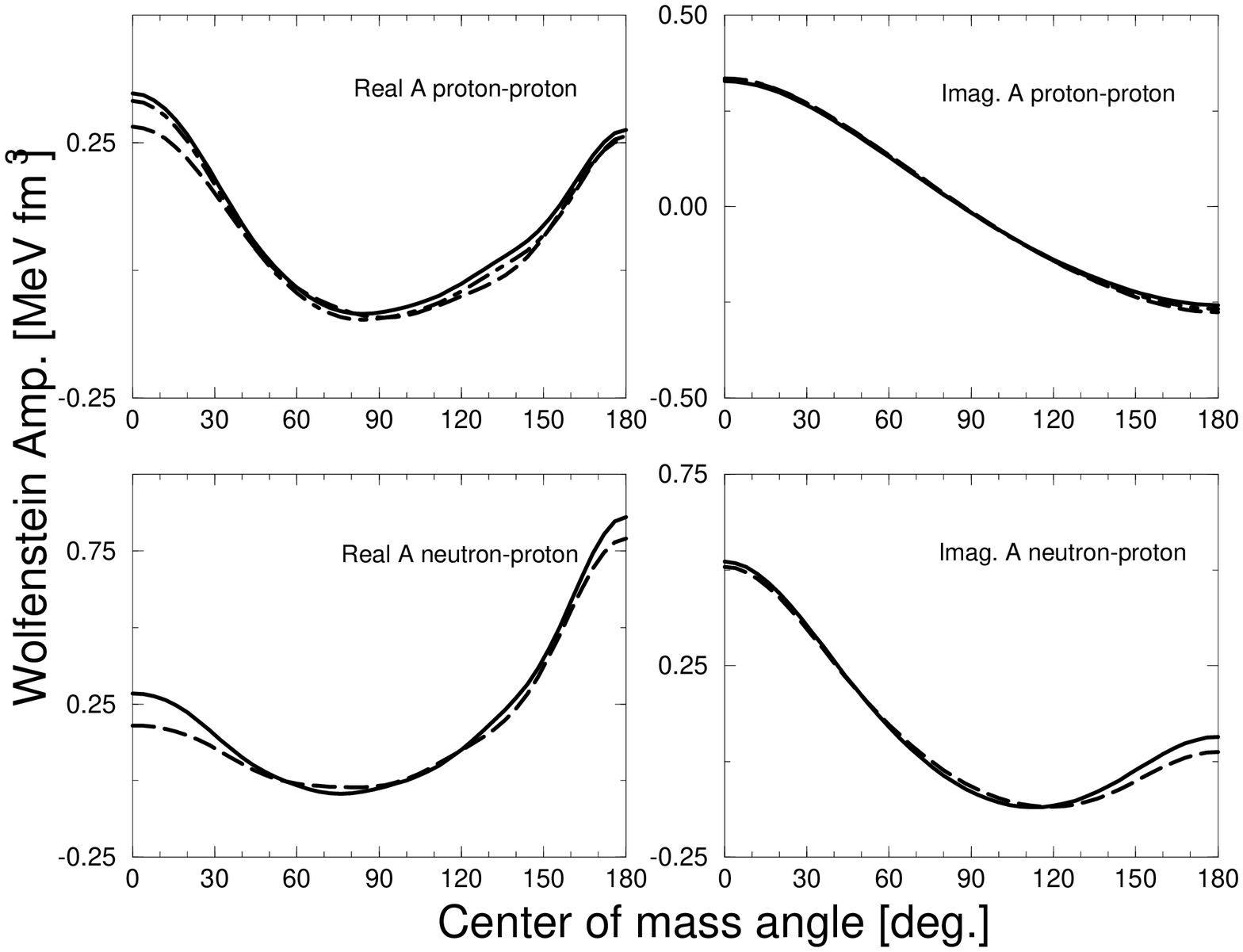}
\caption{
The Wolfenstein amplitude {\it A} for nucleon-nucleon scattering. This
amplitude is a measure of the central piece of the interaction.
The solid line represents the amplitudes fit to the isospin averaged
1999 dataset~\protect\cite{cdbonn2000}. The  dashed line represents
the neutron-proton amplitudes fit to the
1986 dataset~\protect\cite{bonn}. The dash-dot line which only appears
in the proton-proton amplitudes is from the 1999 dataset if
charge dependencies are  used~\protect\cite{cdbonn2000}. Note
that these dependencies are relatively insignificant.
\label{fig3.10}}
\end{figure}
In Fig.~\ref{fig3.10} we also
plot the complex 
central term of the nucleon-nucleon
interaction (Wolfenstein {\it A}~\cite{wolfenstein}) as
a function of angle. The solid
line represents a fit to the 1999 dataset while the dashed
line represents the 1986 dataset. In the real part of
the central term the differences are significant at the extreme
forward angles. These forward angles are kinematically
the most important in the nucleon-nucleus calculation.
Since both the real neutron-proton and real proton-proton
Wolfenstein amplitudes are larger at the extreme forward angle using
the 1999 dataset this shows a direct correlation to the
time dependent growth of
the  
cross section calculations in Figs.~\ref{fig3.1}-\ref{fig3.2}. 
In 1986 the
Watson approximation was a good fit to the experimental data. However
it can be shown that 
by 1993 this was no longer a true statement.

\begin{table*}
\begin{tabular}{||c||c|cccc||}
\hline
\multicolumn{2}{|c|}{total 250 MeV neutron cross section [b]}&
\multicolumn{4}{c||}{Watson Approximation} \\ \hline
target & {\bf EXP.} & {\it 1986}  & {\it 1993}  & {\it 1999}
& {\it 1999CD} \\ \hline
$^{16}$O & {\bf .355}  & .353  & .408  & .403  & .405  \\ \hline
$^{208}$Pb&{\bf 2.786} &2.773  &3.264  &3.224  &3.225  \\ \hline
\multicolumn{2}{|c|}{total 250 MeV neutron cross section [b]}&
\multicolumn{4}{c||}{Many Body Antisymmetry} \\ \hline
target & {\bf EXP.} & {\it 1986}  & {\it 1993}  & {\it 1999}
& {\it 1999CD} \\ \hline
$^{16}$O & {\bf .355}  & .322  & .363  & .357  & .357  \\ \hline
$^{208}$Pb&{\bf 2.786} &2.524  &2.865  &2.818  &2.818  \\ \hline
\end{tabular}
\caption{ The total cross sections of a neutron on $^{16}$O
and $^{208}$Pb. The experimental result is shown (in bold)
as well as a variety
of theoretical calculations based on different nucleon-nucleon
datasets. The datasets are labeled by the year in which they were
defined. The top half of the
table uses only two-body antisymmetrization,
the bottom half of the
table uses many-body antisymmetrization. Error on
experimental and calculation results are significant on the last
digit. The experimental data was taken from
Ref.~\protect\cite{finlay1}. The calculations were produced
as discussed in Fig.~\protect\ref{fig3.1} except that the type of
two-body potential used varied.  The {\it 1986} calculation used the
Bonn potential~\cite{bonn}, the {\it 1993} used the
Nijmegen I neutron-proton averaged potential~\protect\cite{nijmegen}.
Both the {\it 1999} and {\it 1999CD} used the new Bonn potential
described in Ref.~\protect\cite{cdbonn2000}. The CD stands for
charge dependent.
 }
\label{T4.2}
\end{table*}
In Table~\ref{T4.2} neutron total cross section calculations
and experimental data are compared for a 250 MeV neutron impinging
on either $^{16}$O or $^{208}$Pb. The calculations vary by the type
of nucleon-nucleon potential used. The potentials are fit to
different nucleon-nucleon datasets ranging from 1986-1999 and the
calculations use either the traditional Watson approximation (top
half)
or the many-body antisymmetrization (bottom half). The startling
conclusion is that the Watson approximation was the best fit if using
the 1986 dataset but with the introduction of the 1993
dataset the many-body antisymmetrization became the better fit.
With the 1999 dataset the fits of the many-body antisymmetric
technique further improved while the Watson approximation version
became even worse. Again these differences can also
be ascertained directly by examining Figs.~\ref{fig3.8}-\ref{fig3.10}.

The nucleon-nucleus calculation does use the 
off-shell two-body amplitudes. It was shown however that if
two potentials agreed on-shell (they  both fit the nucleon-nucleon
dataset to a high degree of accuracy) than differences off-shell
were near 
inconsequential for nucleon-nucleus elastic
scattering~\cite{tmatrix}. For example in the production of
Table~\ref{T4.2} a Nijmegen potential~\cite{nijmegen}  was used for the
1993 data-set calculation. These results differ with a Bonn
potential~\cite{cdbonn}
calculation
which is based on the same 1993 dataset by at most 1\%.

There are two calculations listed in Table~\ref{T4.2}
that use the 1999 dataset. They
differ on their use of charge dependent terms.  The leftmost of the
two columns (depicted {\it 1999}) uses the neutron-proton
amplitudes for neutron-proton, proton-proton, and neutron-neutron
calculations so their is no charge dependence in this
calculation. Since neutron-proton amplitudes contain both
even and odd momentum space-spin terms it can create a proton-proton
amplitude by using only the odd terms from the neutron-proton
amplitude.
The column labeled {\it 1999CD} calculates
the neutron-proton and proton-proton amplitudes separately based
on different datasets thereby introducing charge dependencies and
thus a more accurate amplitude. These charge
dependencies can not be used in a systematic method for
many-body antisymmetrization.
To calculate the proton-proton
amplitudes of the theory  presented within 
one must use neutron-proton amplitudes
as a source for the even momentum space-spin amplitudes, 
this
process thus nullifies the effect of charge dependencies.
This is why Table~\ref{T4.2} has the same values
for total cross section using the 1999 and 1999CD datasets while
using the many-body antisymmetrization techniques.

Incidentally, the differences in the neutron-proton phase shifts
over time is not without controversy. When the 1993
dataset was defined there were criteria used to refine the set
which threw out
some 
experimental data~\cite{nndata}. There have been concerns
expressed over this procedure's validity and scientific
merit~\cite{cdbonn2000}. This is still an open question and it is 
likely that the nucleon-nucleon dataset will not be static in the
years that follow.

\section{Conclusion}\label{sec3.8}
This work has presented antisymmetry in the
nucleon-nucleus many-body problem.
The new results are a treatment of antisymmetry in
the nucleon-nucleus first-order microscopic
optical potential scattering  calculation with a truer many-body
flavor. The
theoretical  strength of the proton-proton and 
neutron-neutron amplitudes used in the many-body calculations
are cut in half and their character is changed to include states
that have never been used to describe proton-proton scattering
(for example the $^3S_1-^3D_1$ state is now included at half
strength). Traditionally these states 
were referred to as forbidden because they violate the 
Pauli principle in two-body scattering.  These states, although
unphysical in proton-proton and neutron-neutron
scattering in a two-body context,
are physical in the approximate many-body
context presented here.
The many-body
theory also contains exchange in a natural way and allows for
further theoretical development to include inelastic reactions.

Results show that this new many-body antisymmetric theory
represent
the experimental data better  than the Watson approximation
over a large range of reactions. This improvement was only apparent
recently with the use of the newest nucleon-nucleon 
potentials which have significant differences than
their for-bearers.
Many researchers in nucleon-nucleus
scattering theory still use potentials
based on datasets from the 1980's or 
before~\cite{crespo-paris,amos1,neutron} like the
Paris~\cite{paris} or Bonn-B~\cite{bonn} potentials. These potentials
have an extensive history of use in this field and are
still of value if the best effective potentials are sought, but
from a microscopic point of view they are inadequate because
they fail to describe nucleon-nucleon data accurately. 

The use of the microscopic first-order
optical potential in nuclear reaction studies has been
extensive. More fundamental theories have been advanced but it is
still one of the few that is able to produce results when more than a 
few nucleons are involved. This work
has re-examined its antisymmetric character and with little
rigor has suggested a modification which has improved its
validity and power.  Although stronger development is required,
these new contributions will hopefully lend insight
in ascertaining the validity of 
this wonderfully dynamic
scattering theory.

\begin{acknowledgments}
The author would like to
thank the National Partnership for Advanced
Computational Infrastructure (NPACI) under grant No.~ECK200
for the use of their facilities. He would also like to thank 
Ch. Elster as the initial catalyst for this work as well as
H. Arellano, W. Junkin, and S. Karataglidis for helpful discussions
while the work was in progress.
\end{acknowledgments}


\end{document}